\documentclass[epj,referee]{svjour}
\usepackage{graphics}
\begin{document}
%
%\draft
\title{Sensitivity of pQCD deuteron
structure to the nucleon form factors }
\author{Michail P. Rekalo \inst{1}\thanks{\emph{Permanent address:}
\it National Science Center KFTI, 310108 Kharkov, Ukraine}
\and  Egle Tomasi-Gustafsson \inst{2}
}
\institute{Middle East Technical University, 
Physics Department, Ankara 06531, Turkey \and 
DAPNIA/SPhN, CEA/Saclay, 91191 Gif-sur-Yvette Cedex,
France}
\PACS{
{13.40.Gp } {Electromagnetic form factors  } \and
{13.40.-f }{ Electromagnetic processes and properties }
}
\date{Received: date / Revised version: date}
% The correct dates will be entered by Springer
%
\abstract{
We discuss the applicability of  pQCD to the elastic scattering of electrons on protons and deuterons. We analyze the $Q^2$-dependence of the reduced deuteron form factor, taking into account the recent data on the electric proton form factor and we find that the value of the QCD-scale parameter $\Lambda$ differs essentially from the value $\Lambda=0.1$ GeV, previously found using the dipole parametrization of the electromagnetic nucleon form factors $G_E$ and $G_M$. Moreover the predicted scaling behavior of the reduced deuteron form factor can not be recovered in the Dirac and Pauli representations for the nucleon electromagnetic form factors. 
}
\maketitle

\section{Introduction}
The internal structure of hadrons is conveniently described in terms of electromagnetic form factors (FFs). FFs are on one side, experimentally accessible (through cross section and polarization observables measurements) and on the other side, appear explicitely in the expression of the electromagnetic current allowing a straightforward comparison with theoretical models.
The traditional way to measure hadron electromagnetic FFs is through  elastic scattering of electrons.
In the intermediate region of momentum transfer squared 
one of the the main physical issues, related to the study of the 
deuteron structure, is to 
determine the kinematical region where the transition to pQCD occurs, i.e. 
where a description in terms of quark and gluon degrees of freedom would be  more adequate than  a picture of mesons and nucleons.  

The authors of \cite{Al99} suggested that the data about the structure
function $A(Q^2)$ in $ed$ elastic scattering, in the range of momentum transfer
$2\le Q^2\le$ 6 GeV$^2$,  are a good indication of the validity of the predictions of pQCD. More
exactly, following \cite{Br76}, let us introduce a generalized deuteron FF, $F_D(Q^2)$, $F_D(Q^2)=\sqrt{A(Q^2)}$, and a  reduced deuteron FF $f_D(Q^2)$:
\begin{equation}
f_D(Q^2)=\displaystyle\frac{F_D(Q^2)}{F_N^2(Q^2/4)},
\label{eq:eq1}
\end{equation}
where $F_N$ is the nucleon electromagnetic FF. The $Q^2$-behavior of 
$f_D(Q^2)$ (at large $Q^2$) can be predicted in the framework of pQCD, in the following form:
\begin{equation}
f_D(Q^2)=N\displaystyle\frac{\alpha_s(Q^2)}{Q^2}\left ( ln \displaystyle\frac{Q^2}{\Lambda^2}\right)^{-\Gamma},
\label{eq:eq2}
\end{equation}
where N is the normalization factor (which is not predicted by QCD), $\alpha_s$ is the running QCD strong interaction coupling constant, taken here as $\alpha_s=\left [ ln ({Q^2}/{\Lambda^2})\right]^{-1}$ following refs. \cite{Al99,Br76}, $\Lambda$ is the scale QCD parameter, and $\Gamma$ is determined by the leading anomalous dimension, here $\Gamma=-8/145$. 

In \cite{Al99} it was shown that the QCD prediction (\ref{eq:eq2}), which can be applied to asymptotic momentum transfer, is working well already for $Q^2\ge 2$ GeV$^2$, with a plausible value of the parameter
$\Lambda \simeq$ 100 MeV, in agreement with the values determined by many other possible methods \cite{pdg}.

We note that in  \cite{Br76} another interesting prediction, concerning the scaling behavior of the reduced deuteron FF, was done:
\begin{equation}
f_R=\left ( 1+\displaystyle\frac{Q^2}{m_0^2}\right )f_D(Q^2)\simeq const,
\label{eq:eq3}
\end{equation}
where $m_0^2=0.28$ GeV$^2$ is a parameter related to the $Q^2$-behavior of the  pion FF. The same data from \cite{Al99}, if plotted in the representation of the reduced deuteron FFs, should illustrate the $Q^2$-independence of this product.

\begin{figure*}
\resizebox{0.75\textwidth}{!}{%
  \includegraphics{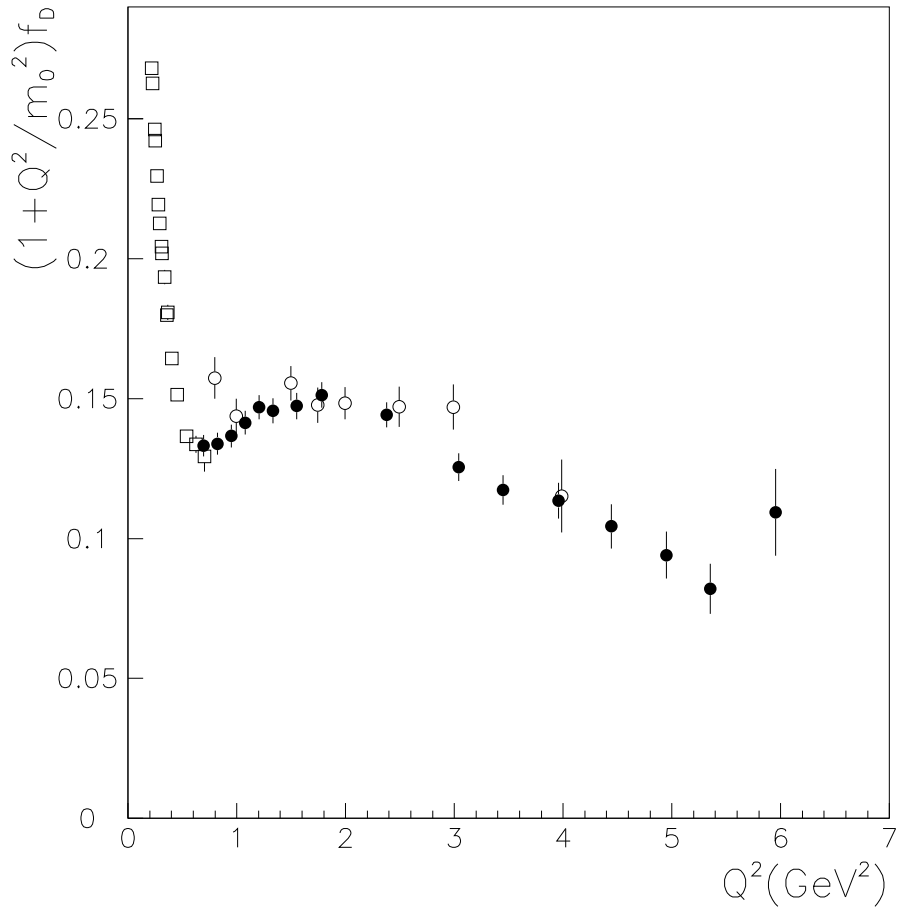}}
\vspace*{2truecm}
\caption{Data set corresponding to the reduced deuteron FFs multiplied by $(1+Q^2/m_0^2)$. Open circles are from \protect\cite{Ar86}, open squares from \protect\cite{Pl90}, solid circles from \protect\cite{Al99}.}
\label{fig:fig0}
\end{figure*}

This result was confirmed by the previous $A(Q^2)$ data \cite{Ar86}, in the limit of their accuracy. In Fig. (\ref{fig:fig0}) we  show  that the new, more precise data about $A(Q^2)$ \cite{Al99}, are not consistent with the prediction (\ref{eq:eq3}) as they show an evident dependence of the product $f_R$ on $Q^2$. This behavior can not be changed by varying the parameter $m_0$. Therefore a contradiction appears in the interpretation of hadron electromagnetic FFs (nucleon and deuteron) in terms of pQCD, in the intermediate $Q^2$-range (at least up to $Q^2=$ 6 GeV$^2$) and this motivated us to analyze the predictions (\ref{eq:eq2}) and (\ref{eq:eq3}) in more detail, with respect to the value of the parameter $\Lambda$ and, in particular, to the choice of the nucleon FFs in Eq. (\ref{eq:eq1}).

\section{Analysis of deuteron and nucleon form factors}

Let us discuss firstly the definition of the reduced deuteron FF, following Eq. (\ref{eq:eq1}). More exactly, the reduced deuteron form factor, in ref. \cite{Br76}, has been defined as $f_D(Q^2)=F_D(Q^2)/[F_p(Q^2/4)F_n(Q^2/4)]$, where $F_p$ and $F_n$ are the nucleon FFs. But which ones? A 'generalized nucleon FF', $F_N$ was parametrized in dipole form: 
\begin{equation}
F_N(Q^2)= G_D=\displaystyle\frac{1}{(1+Q^2/m_D^2)^2}, ~m_D^2=0.71\mbox{~GeV}^2, 
\label{eq:eq4}
\end{equation}
and it was not rigorously identified as magnetic or electric. The proton and neutron FF were assumed to be equal.

Note in this respect, that the dipole form of the nucleon electromagnetic FFs has been taken until recently as universal, consistent with the experimental data for three of the four nucleon FFs, $G_{Mn}$, $G_{Mp}$, and $G_{Ep}$:
$$G_{Ep}(q^2)=G_{Mp}(q^2)/\mu_p=G_{Mn}(q^2)/\mu_n=G_D,$$
$$\mu_p=2.79,~\mu_n=-1.91.$$
The $(1/Q^2)^2$-behavior of these FFs are in agreement with quark counting rules considerations \cite{Br76}. The fourth FF, $G_{En}$, was assumed negligible in the discussed region of $Q^2$.

\begin{figure*}
\resizebox{0.75\textwidth}{!}{%
  \includegraphics{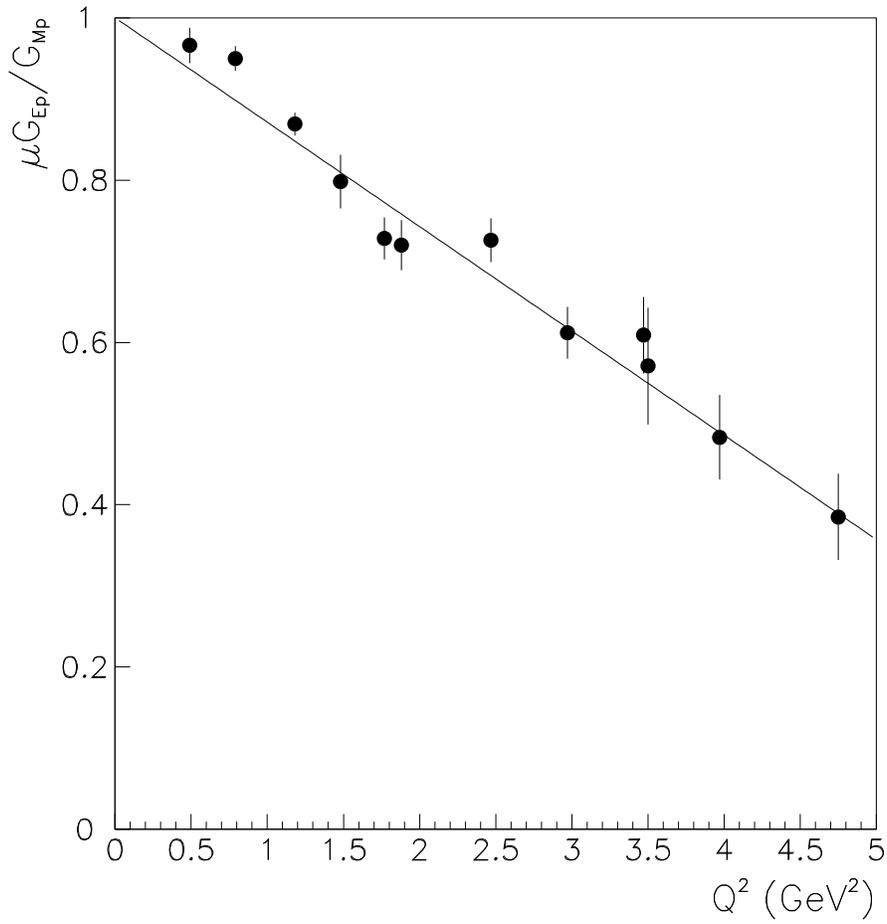}}
\vspace*{2truecm}
\caption{Proton electric FF, $G_{Ep}$ (normalized to $G_{Mp}/\mu\simeq G_D$),  from \protect\cite{Jo00} (circles). The solid line is the parametrization according to Eq. (5). }
\label{fig:gep}
\end{figure*}

The experimental data about elastic $eN$-scattering, based on the Rosenbluth separation of the two possible contributions to the differential cross section with unpolarized particles \cite{Ar86,And94} were consistent with this representation, until when, recent more precise data \cite{Jo00}, based on the  polarization method \cite{Re68}, showed that the $Q^2$-dependence of the nucleon electromagnetic FFs is not universal, and that the electric proton FF strongly deviates from the usual representation. 
We will use, for the description of the data \cite{Jo00}, a fit of the form: 
\begin{equation}
G_{Ep}(Q^2)= \displaystyle\frac{1}{(1+Q^2/m_D^2)^2}{(1-0.129*Q^2)},
\label{eq:eq5}
\end{equation}
where the second factor explicitly shows the deviation from the dipole form
(solid line in Fig. \ref{fig:gep}).
These data can also be described by a dipole form, but changing the mass parameter 
$m^2_D=0.71$ GeV$^2$ to a smaller value: $m^2_D=0.6$ GeV$^2$. Such  parametrization may seem preferable, because it is consistent with the pQCD counting rules, but the new best fit value of $m^2_D$ is in contradiction with the nice relation between the $Q^2$-behavior of pion and nucleon FFs, derived in \cite{Br76}. Note that a similar deviation from the dipole fit has also been recently observed for the $N\to\Delta$ transition FF \cite{Fr99}. Concerning the FF $G_{En}$, a recent anaysis of all electron-deuteron elastic scattering observables \cite{Etg01} shows that, even in the framework of the standard non relativistic impulse approximation \cite{BJ76}, one can get a good general description, with $G_{En}$ in agreement with the QCD-inspired parametrization \cite{G-K}. More direct measurements have been recently done, for $Q^2\le 2$ GeV$^2$, showing definitely that $G_{En}$ does not vanish \cite{Zh01}.

So from the recent experimental data on $eN$-elastic scattering, the $Q^2$-behavior of nucleon electromagnetic FFs is consistent with:
\begin{itemize}
\item "standard" dipole function for the nucleon magnetic FFs $G_{Mp}$ and $G_{Mn}$,
\item linear deviation from the dipole function for the electric proton FF $G_{Ep}$,
\item non vanishing electric neutron FF, $G_{En}$.
\end{itemize}

The important question is, then, which parametrization of nucleon FFs to use in calculating the reduced deuteron FF $f_D(Q^2)$, Eq. (1), and what are the consequences of different choices on the apparent value of the parameter $\Lambda$. 

\begin{figure*}
\resizebox{0.75\textwidth}{!}{%
  \includegraphics{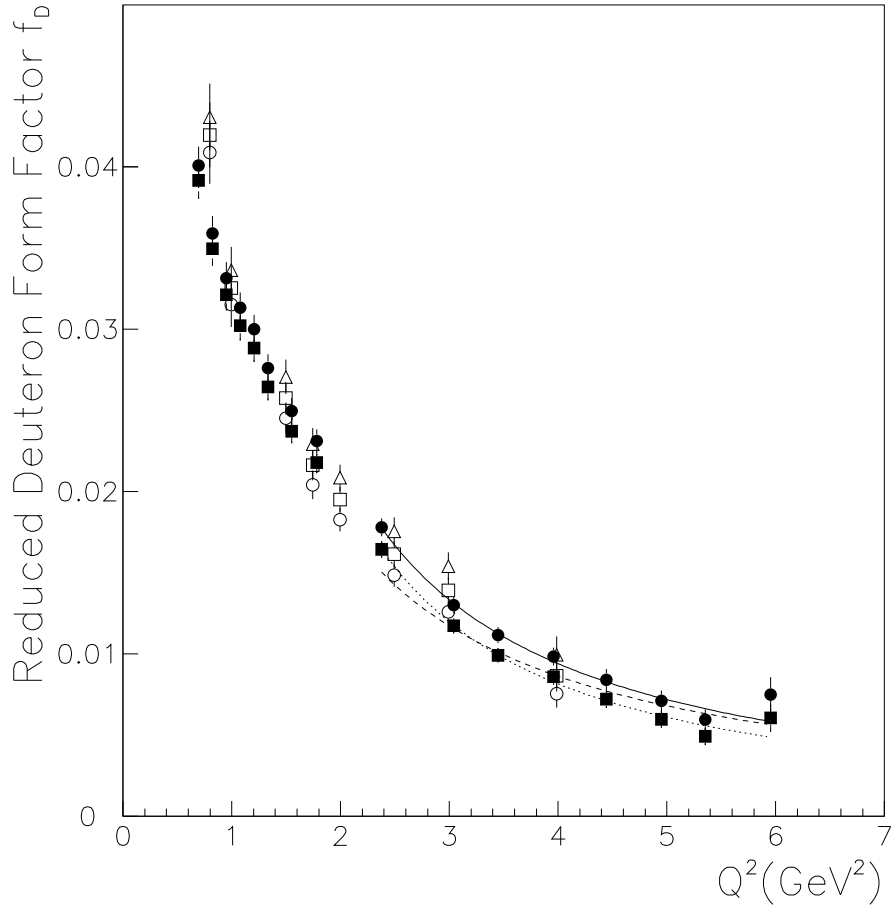}}
\vspace*{2truecm}
\caption{Data set corresponding to the reduced deuteron FF for different choices of the generalized nucleon FF: circles (case 1), squares (case 2), and triangles (case 3). Open symbols are from \protect\cite{Ar86}, solid symbols from \protect\cite{Al99}.}
\label{fig:fig1}
\end{figure*}

In Fig. (\ref{fig:fig1}) we show different data sets and best fits, using Eq. (\ref{eq:eq2}), corresponding to the following possibilities:
\begin{enumerate}
\item We replace in Eq. (\ref{eq:eq1}) $F_N(Q^2/4)$ by the  fit (\ref{eq:eq5}) of new data on the proton electric FF, $G_{Ep}$: $$f_D(Q^2)=\displaystyle\frac{F_D(Q^2)}{G_{Ep}^2(Q^2/4)}.$$ This yields to the data set represented by circles and to the fit reported as a solid line (case 1).
\item  We replace in Eq. (\ref{eq:eq1}) $F_N^2(Q^2/4)$ by the product of $F_N$ (Eq. 4) and $G_{Ep}$ from Eq. (\ref{eq:eq5}): $$f_D(Q^2)=\displaystyle\frac{F_D(Q^2)}{F_N(Q^2/4) G_{Ep}(Q^2/4)}.$$ The $f_D$ data are shown as squares and the best fit by the dashed line (case 2).
\item We show, for comparison, the previous results of Ref. \cite{Al99}, using the dipole parametrization Eq. (\ref{eq:eq4}). The data are represented by triangles and the fit by the  dotted line (case 3).
\end{enumerate}

In all these three cases, instead of normalizing the QCD prediction, Eq.  (\ref{eq:eq2}), to the data at $Q^2= 4$ GeV$^2$, as in Ref. \cite{Al99}, we have fitted the data beyond $Q^2= 2$ GeV$^2$,  with two free parameters, a global normalization $N$ and $\Lambda$.  We found that even a relatively small change in nucleon FFs, causes a relatively large instability in the value of $\Lambda$. Note that the reduced FF $f_D$ has logarythmic (i.e. relatively weak) dependence on the parameter $\Lambda$, Eq. (2).  For the case 3, we obtain a different value for the parameter $\Lambda$, as compared with Ref. \cite{Al99}. This is due to the different normalization procedure. 
A similar situation occurs if we use the Dirac and Pauli FFs, 
\begin{equation}
F_1=(G_E+\tau G_M)/(1+\tau) \mbox{~and~} F_2=(G_M-G_E)/(1+\tau),
\label{eq:eq6}
\end{equation}
(where $\tau=Q^2/(4M^2)$, $M$ is the nucleon mass), instead of the Sachs FFs $G_E$ and $G_M$. In principle, both sets of nucleon FFs correspond to an equivalent description of the electromagnetic structure of the nucleon, but the physical properties associated to these two sets of FFs are different. For example, their asymptotic behavior is different: for example, if the Sachs FFs satisfy the  the Phragm\`en-Lindel\"of  theorem   \cite{Ti39} 
(concerning the equality of these FFs at large values of space- and time-like momentum transfer squared \cite{Re01}), then the asymptotic behavior of the FFs $F_1$ and $F_2$ will not satisfy the above mentioned theorem. Note that, from a theoretical point of view, the FFs $F_1$ and $F_2$ seem more fundamental, as they enter into a parametrization of the electromagnetic current in a relativistic and gauge-invariant form, valid in any coordinate system. On the other hand, the Sachs electromagnetic FFs can be related to the distributions of the electric charge and magnetic moment of the nucleon only in the Breit system. Moreover, the analytical properties of the FFs have been studied in terms of $F_1$ and $F_2$  \cite{Dr61}. 

So let us substitute in Eq. (1) the nucleon FF $F_N^2$ by different combinations of $F_1$ and $F_2$. We calculate $F_1$ and $F_2$ for the proton,  with the expression (\ref{eq:eq5}) which fits the new $G_{Ep}$ data and use the dipole form for $G_{M}$. 

\begin{figure*}

\resizebox{0.75\textwidth}{!}{%
  \includegraphics{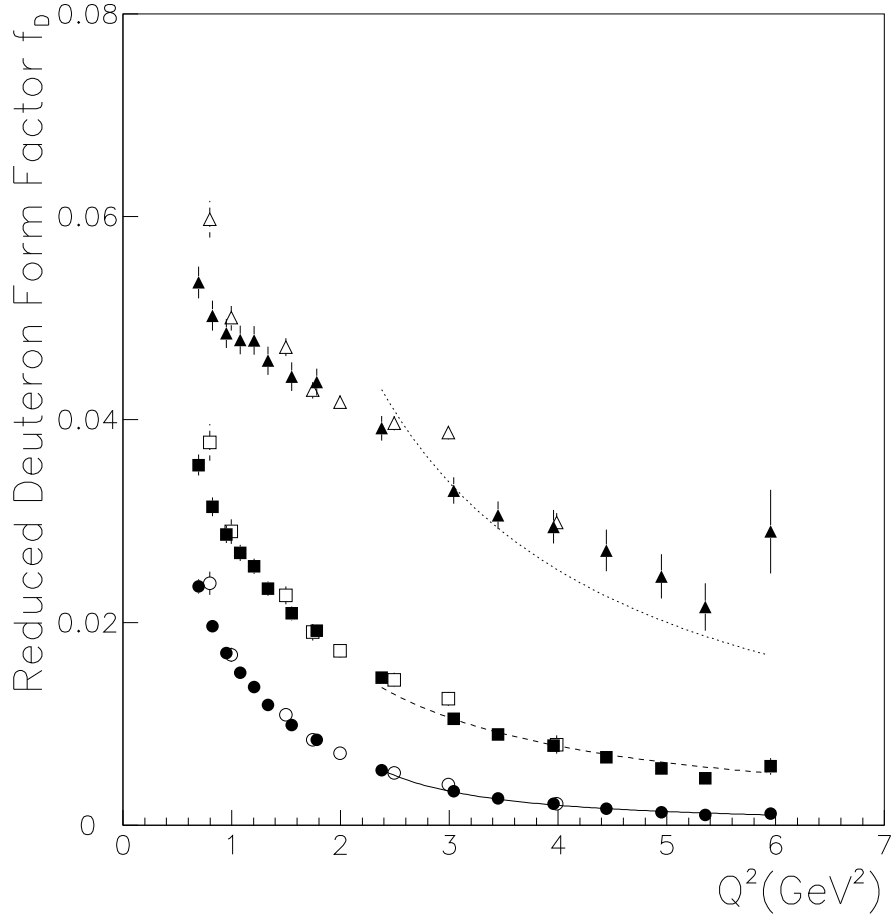}}
\vspace*{2truecm}
\caption{Data set corresponding to the reduced deuteron FF for different choices of the nucleon FFs, in terms of $F_1$ and $F_2$. circles(case 4), squares (case 5), and triangles  (case 6). Open symbols are from \protect\cite{Ar86}, solid symbols from \protect\cite{Al99}.}
\label{fig:fig2}
\end{figure*}

In Fig. \ref{fig:fig2}  we illustrate the following calculations:

\begin{enumerate} 

\item  We replace in Eq. (\ref{eq:eq1}) $F_N^2(Q^2/4)$ by  $F_1^2(Q^2/4)$ (circles). The best fit is shown as solid line (case 4);

\item  We replace in Eq. (\ref{eq:eq1}) $F_N^2(Q^2/4)$ by the product of $F_1(Q^2/4)$ and $F_2(Q^2/4)$   (squares). The best fit is shown as dashed line (case 5);

\item We replace in Eq. (\ref{eq:eq1}) $F_N^2(Q^2/4)$ by  $F_2^2(Q^2/4)$  (triangles). The best fit is shown as dotted  line (case 6);
\end{enumerate}

The values of the fitting parameters $N$ and $\Lambda$ are summarized in Table 1. The values which can be obtained for $\Lambda$ may differ of orders of magnitudes, for the different possible choices of the nucleon FFs for the calculation of the reduced FF $f_D(Q^2)$. The  normalization parameter $N$ also shows large sensitivity to the choice of the nucleon FF. One can see, from this Table, that a physically acceptable value of $\Lambda\simeq 200$ MeV, appears only for a specific choice, $F_N^2=G_{Ep}^2$ (case 1). But the dipole choice, coherent with QCD, (case 3), or the helicity conserving choice, (case 4), result in too large values for the parameter $\Lambda$. 

\begin{figure*}
\resizebox{0.75\textwidth}{!}{%
  \includegraphics{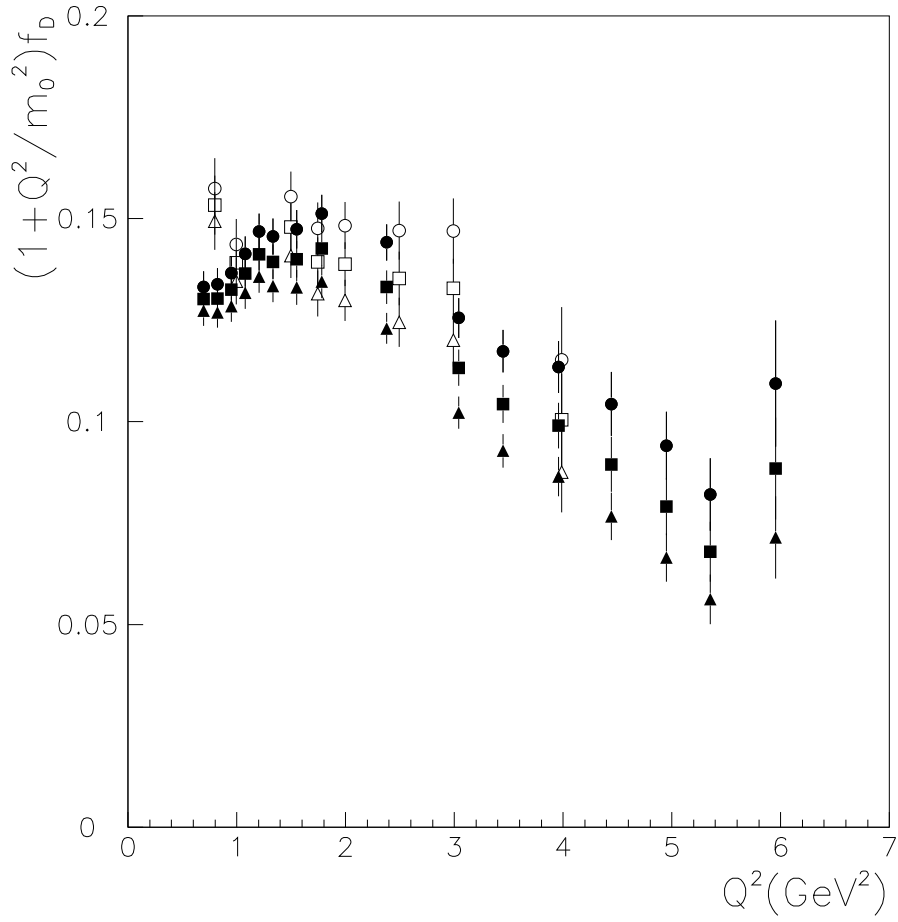}}
\vspace*{2truecm}
\caption{Data set corresponding to the reduced deuteron FFs multiplied by $(1+Q^2/m_0^2)$. Notations as in Fig. 3, $m_0^2$=0.28 GeV$^2$.}
\label{fig:fig3}
\end{figure*}

The reduced deuteron FFs, according to any of these choices, violate essentially the $Q^2$-independence of the product: $ \left (1+\displaystyle\frac{Q^2}{m_0^2}\right )f_D(Q^2)\simeq const$
which has been pointed out earlier \cite{Br76}. This is true for the different choices of electromagnetic  nucleon FF (Fig. \ref{fig:fig3}). This result is also quite insensitive to different values of the $m_0$ parameter.

One should also take into account the fact that the elastic $ed$-scattering is sensitive to the isoscalar combination of the nucleon FFs $G_{Es}$ and $G_{Ms}$, with $2G_{Es}=G_{Ep}+G_{En}$ and $2G_{Ms}=G_{Mp}+G_{Mn}$. So the corresponding linear combination of proton and neutron FFs seems more adequate for the parametrization of $F_N$. In the case of $G_{En}=0$ and dipole parametrization of the other nucleon electromagnetic FFs, an isoscalar combination will only bring a different normalization. But, if one takes $G_{Ep}\ne G_D$ and $G_{En}\ne$ 0 (as indicated by the present experiments), two other possibilities: $F_N^2=G_{Es}^2$ and $F_N^2=G_{Es}G_{Ms}$ would lead to different results and different values for the parameter $\Lambda$.

Up to now we studied the sensitivity of the reduced deuteron FF to different choices of nucleon FFs. However, the numerator of Eq. (\ref{eq:eq1}) contains a generalized deuteron FF, derived from the structure function $A(Q^2)$, which is a quadratic function of the three deuteron electromagnetic FFs. It would be more natural to include the electric, quadrupole or magnetic deuteron FFs, $G_E$, $G_Q$, and $G_M$ in the calculation of $f_D$, more exactly, in the helicity conserving FF. The deuteron magnetic FF has been measured, (through the SF $B(Q^2)$) up to $Q^2=2.77$ GeV$^2$ \cite{Bo90}. The separation of the electric and quadrupole deuteron FFs can be done only after the measurement of a tensor polarization observable in $ed$-elastic scattering. A dedicated experiment at JLab \cite{Ab00} has shown that this is possible up to  $Q^2=1.7$ GeV$^2$. At larger $Q^2$, polarization measurements in $ed$-elastic scattering become very difficult, due to the steep decreasing of the differential cross section with $Q^2$. Note that, the considered region of $Q^2$ contains the points of zero crossing of  the charge FF $G_C$ (at $Q^2\simeq 0.7$ GeV$^2$) and of  the magnetic FF $G_M$ (at $Q^2\simeq  1.8$ GeV$^2$. In this respect the quadrupole FF $G_Q$ would be more convenient, as a starting point for the calculation of the reduced deuteron FF, but no experimental point exists over $Q^2=1.7$ GeV$^2$.

\section{Conclusions}

We have shown that the present situation with nucleon and deuteron electromagnetic FFs, at the light of the recent $G_{Ep}$ data in the intermediate $Q^2$ range, is less clear than in case of dipole parametrization for the proton electromagnetic FFs:  the scaling laws predicted by QCD do not apply to the deuteron FFs in
the intermediate $Q^2$ range, where precise data exist on the elastic
electron deuteron cross section.  We have used different possible
parametrizations for the nucleon FFs and shown that the overall results
are not consistent with the pQCD predictions: the results of \cite{Jo00} open the way to new interpretations in different directions. We must stress that our work is essentially  based on the existing experimental information about nucleon FFs.

The last experimental data about the differential cross sections for other deuteron electromagnetic processes, $\gamma +d\to d+\pi^0$ and  $\gamma +d\to n+p$ also show a deviation from the QCD predictions concerning the reduced matrix elements. A new reformulation of this problem, in case of pion photoproduction has been recently done \cite{Br02}, assuming the scaling of the deuteron FFs. However, the predictions for both these processes,  $\gamma +d\to d+\pi^0$ and  $\gamma +d\to n+p$ will also strongly depend on the choice of the nucleon electromagnetic FFs in the definition of the reduced nuclear matrix element. 

We thank C. Glashausser for a critical reading of the manuscript and useful remarks. We thank S. Brodsky for bringing to our attention Ref. \cite{Br02}.

\clearpage\
%%%%%%%%%%%%%%%%%%%%
% Table
%%%%%%%%%%%%%%%%%%%%%

\begin{table}
\caption{ Values of the fit parameters, see text.}
\label{tab1}
\begin{tabular}{|l|c|c|c|c|}
\hline\hline
Case&N & $\Lambda [GeV]$&$\chi^2$\\
\hline\hline
 (1) $F_N^2= G_{Ep}^2$& 0.16 $\pm$ 0.04& 0.20  $\pm$ 0.13& 0.7\\
 (2) $F_N^2= G_{Ep}G_{Mp}$&0.43 $\pm$ 0.04& 0.0014$\pm$ 0.007& 3.2 \\
 (3)$ F_N^2=G_{Mp}^2$&0.06 $\pm$ 0.02 & 0.6$\pm$ 0.2& 0.7\\  
 (4) $F_N^2= F_1^2$& 0.01 $\pm$ 0.001& 1.1  $\pm$ 0.1&1.0 \\
 (5) $F_N^2= F_1F_2$&0.43 $\pm$ 0.04& 0.006$\pm$ 0.004&2.0 \\
 (6) $F_N^2= F_2^2$&2.9 $\pm$ 1.00 & 0$\pm$ 4& 5.9\\
 \hline\hline
\end{tabular}
\end{table}

\end{document}